# Sports highlights generation based on acoustic events detection: A rugby case study

Anant Baijal, Jaeyoun Cho, Woojung Lee and Byeong-Seob Ko

# Sports highlights generation based on acoustic events detection: A rugby case study

*Anant Baijal, Jaeyoun Cho, Woojung Lee and Byeong-Seob Ko*
Multimedia R&D Team, DMC R&D Center
SAMSUNG Electronics Co., Ltd., Suwon, South Korea 443-742
{baijal.anant, jyoun.cho, w0320.lee, b.ko}@samsung.com

*Abstract*—We approach the challenging problem of generating highlights from sports broadcasts utilizing audio information only. A language-independent, multi-stage classification approach is employed for detection of key acoustic events which then act as a platform for summarization of highlight scenes. Objective results and human experience indicate that our system is highly efficient.

## I. INTRODUCTION

The field of sports highlights detection or sports video summarization is extensively researched. Prior studies have used audio cues [1-5], visual cues [6], mixture of audio-visual cues [7-8], audio-textual cues [9], and a variety of features and classifiers to generate highlights [10]. Studies have also used textual cues from social media for generating sports highlights [11]. For a task like automatic highlights generation using live TV broadcasts (or stored sports broadcasts), audio cues are a natural choice owing to their light computational needs and their feasibility in confirming to real-time requirements.

Like sports themselves, the audio contents of each sport have their own unique characteristics: the game of rugby can be characterized by players' shouts, applause, whistles etc.; a basketball game can have dribbling sounds, sounds of ball hitting the board or sounds of players screeching the wooden court. Crowd noise and commentators' speech are common sounds found in many sports broadcasts. By analyzing certain audio characteristics, one can automatically generate highlights for sports.

Some studies in the past have applied their approach or features for detecting highlights in more than one sport [5], but it may not be efficient to do so, given that some of the sounds found in one sport may be absent in another, and furthermore, the same sounds in different kind of sports can have different characteristics.

Not much research has targeted the extremely popular sport of rugby. Based on our analysis of the audio contents of a number of rugby matches, we design the training mechanism and highlights generator engine, which are presented in Sec. II of this paper. Our approach can be utilized to mine highlight scenes from both live broadcasts and stored sports content.

Audio based content analysis of rugby offers the following challenges: i) scoring events like 'try' do not produce distinct sounds (unlike ball-hit in baseball or kick in soccer); ii) the broadcast content is corrupted by random noises like music, shouts etc.; iii) the crowd is in 'cheer' mode for most part of the sport; iv) the broadcasts of matches differ in acoustic settings and properties.

The focus of our work is twofold: i) acoustic events detection with high recall rate and negligible precision error; ii) generation of highlight scenes using detected events.

The organization of the paper is as follows: our system comprising of training stage and highlights generation engine is presented in Section II. The dataset and both objective and subjective performance results are highlighted in Section III. Discussion and conclusions are presented in Section IV.

## II. SYSTEM

### A. Training stage

The training mechanism has three modules as shown in Fig.1, and explained in the following sub-sub sections:

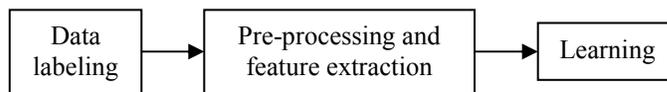

Fig. 1. Training Stage

*1) Data Labeling*

Based on audio contents of rugby matches, we observe the following acoustic events as representative candidates to generate highlights:

● *Referee's whistle*

A referee's whistle is one the most important indicators of when a scene could be classified as highlight as this acoustic event happens in case of *try*, severe fouls, conversion kicks and other important events. Unlike many sports such as soccer, basketball or ice hockey, the sound of whistle in rugby broadcasts is very prominent (see Fig.2) and clearly audible in the audio stream of the broadcast. The audio contents of rugby matches have a whistle sound that is distinctive- it has a shrill tone and a pitch high enough for the whistle to be heard despite high crowd noise or commentators' speech.

● *Commentators' excited speech*

We observe that the commentators' speech turns excited during a highlight scene while in other scenes it is unexcited and hence there is a strong correlation between an exciting scene and a commentators' speech. It is to be noted that the commentators' speech is corrupted with other sounds during the match including crowd noise.

● *Crowd Noise*

A number of previously cited studies have used crowd noise as a cue for various sports. In rugby, however, we observe that the crowd, for most part, is in a high cheering - highly excited-highly ebullient state, and so is the crowd noise. Acoustically, there is not much variation in the crowd noise in presence or absence of a highlight scene, and thus although a representative candidate, this category of sound is *not* used as a deciding event for highlights generation.

Thus only commentators' *excited speech* and referee's *whistle sounds* are considered as deciding events for audio based highlights in rugby. Additionally, we also label events like unexcited speech and music.

*2) Pre-processing and feature extraction*

The input audio signal is divided into *frames* and converted into spectral domain. We then employ Mel Frequency Cepstral Coefficients (MFCC) [12] and their first order differential coefficients, commonly known as delta-MFCC, to extract requisite features.

The combination of MFCC and delta-MFCC works out very well for representing both the whistle sounds and the change in excitement level of speech: while MFCCs can capture the power spectral envelope of a signal and have proven to be very efficient for speech analysis, delta-MFCCs capture the dynamics, i.e., the variation of MFCC coefficients with time.

To tackle the problem of amplitude variation, we ignore the zeroth order coefficient of MFCC vector. In order to increase the detection rate of acoustic events, the MFCC and delta-MFCC features are extracted from the incoming audio stream but at different frame intervals. The feature vector used is 60-dimensional in total.

Some spectrograms of typical audio events taken from actual rugby broadcasts are shown in Fig.2. It can be seen that the spectral contents of whistle, excited speech, and overlap of the two, clearly differ from each other.

*3) Learning*

We make use of Gaussian Mixture Models (GMM) to learn the extracted features. GMMs are composed of probabilistic models, where each model has a mean and a co-variance matrix associated with it.

$$p(c|\lambda) = \sum_{i=1}^{M} w_i \cdot g(c|\mu_i, \Sigma_i) \quad (1)$$

Here,

$p(c|\lambda)$ represents a mixture model,

$c$ represents the feature vector,

$w_i$; $i = 1,2,3......M$ denotes the mixture weights

$g(c|\mu_i, \Sigma_i)$ represents the conditional Gaussian densities with mean $\mu_i$ and covariance matrix $\Sigma_i$.

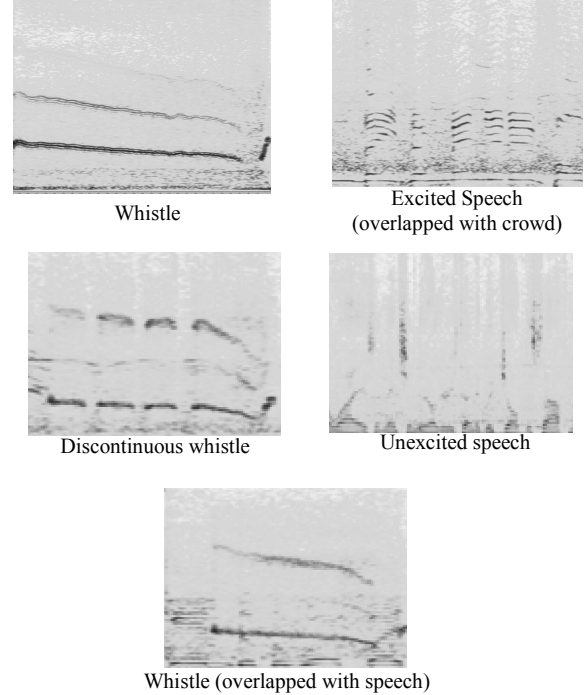

Whistle

Excited Speech (overlapped with crowd)

Discontinuous whistle

Unexcited speech

Whistle (overlapped with speech)

Fig. 2. Spectrograms from actual broadcasts

*B. Highlights Generator Engine*

Based on our observations and analysis presented in the above sub-section, we present an algorithm that uses computationally inexpensive features combined with a simple yet efficient approach, to generate highlight scenes from rugby broadcasts. The algorithm can be seen in Fig. 3 and individual modules are explained in the following sub- subsections.

*1) Pre-processing and feature extraction*

The features are extracted in the same manner as explained earlier. It is to be noted that high noise content (crowd or music or other sounds) is present in the background in addition to the commentators' speech and referees' whistle sounds.

*2) Multi-stage classification*

It can be observed from Fig.2 that there is an overlap of other sounds (unexcited speech, crowd etc.) with key acoustic sounds, hence it becomes challenging for the classifier to categorize the events correctly as the learning models have only a limited discerning ability in case of a single-stage classification. For e.g., a segment having dominating 'crowd noise' overlapping with 'excited speech' may be classified under 'crowd' category, implying that we missed detecting a key event (excited speech). Multi-stage GMM classification merits in providing *preferential classifying power* (ability to choose the preferred audio event category in case of an overlap), thereby directly increasing the recall rate of key acoustic events.

In our approach, the trained GMM models first classify the incoming audio frames as either 'speech' or 'non-speech' acoustic events as the feature vector used is extremely efficient in detecting speech segments. This stage results in a high detection rate for 'excited speech' segments even in presence of overlapping non-speech sounds. The audio segments classified as 'speech' in the first stage are then passed through a second stage of classification after which they are classified by the GMM as 'excited speech' or 'unexcited speech' while the non-speech segments are categorized as either 'whistle' or 'others'. We thus train five GMM models- a speech model, an excited speech model, an unexcited speech model, a whistle model and a model to classify all other acoustic events. The occurrences of 'excited speech' and 'whistle' are continuously stored in a buffer along with their timestamp information.

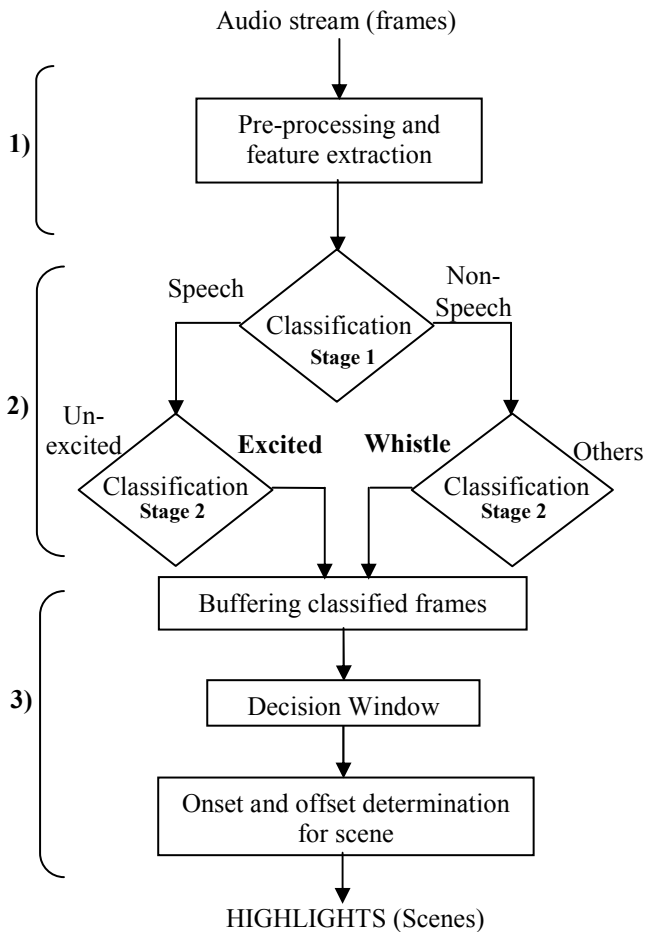

Fig. 3. Highlights Generator Engine for Rugby
1) Pre-processing and feature extraction 2) Multi-stage classification
3) Post-processing

*3) Post-processing*

A decision *window* of a certain duration is glided over the *classified frames* that are continuously buffered. Now if the percentage of events of interest ('whistle' and 'excited speech') in the window exceeds a set threshold, the starting point of that decision window (the timestamp of the first frame in the window) becomes the onset of the highlight scene. To determine the endpoint of the highlight scene, the decision window's starting (and thus, the ending point) are shifted by one until the shifted window also meets the set percentage threshold, and when it does not, the endpoint of the highlight is fixed as the endpoint of the previous window (timestamp of the last frame in the previous window). This process is then repeated, the starting point being the immediate next frame to the endpoint of the highlight scene. The reasons for not generating highlight as and when each window meets the threshold condition are twofold: i) the presence of commentators' excited speech and whistle is strongly correlated with a scene being a highlight but these acoustic events may occur anywhere during the scene, even towards the ending of the scene; ii) this would imply that each highlight is of the same duration (which is not the case in actual sports).

### III. EXPERIMENTAL RESULTS

*A. Dataset*

We use actual broadcasts of a wide variety of rugby matches. The audio stream from rugby matches is converted to 16 kHz, 16-bit, mono-channel format. The training data are manually annotated for 11 different games to determine the ground truth for various audio event categories. For evaluation purposes, our test dataset consists of 5 games totaling 10 hours and 30 minutes in duration.

*B. Objective evaluation*

We first calculate the recall and precision rates for detecting key acoustic events in both the classification stages (see Fig.3) as the generation of highlights is directly dependent on them. For classification stage 1, we pick 2808 'speech' segments, while for classification stage 2, we pick 217 'excited speech' segments and 112 'whistle' segments (other than training data) and test our approach; the segments are of different duration. The results in Table I show that our approach yields extremely high recall-precision rates in both the stages.

TABLE I
PRECISION-RECALL RATES FOR ACOUSTIC EVENTS DETECTION IN RUGBY

| Classification Stage | Event Category | # of test instances | Recall (%) | Precision[^] (%) |
|---|---|---|---|---|
| Stage 1 | Speech | 2808 | 97.19 | 96.49 |
| Stage 2 | Excited Speech | 217 | 96.77 | 97.37 |
|  | Whistle[*] | 112 | 97.32 | 99.72 |

[*]The 'whistle' segments also involve cases where there is an overlap of whistle and speech sounds.
[^] The precision is calculated by testing with other acoustic event categories such as crowd, music, unexcited speech and miscellaneous sounds.

Next, we evaluate the overall performance of our approach for detecting highlights. The objective results in Table II indicate that our approach to detecting highlights in rugby yields promising results, with an overall *try* recall rate of 97.06% with a 93.41 % precision for highlight scenes. A consumer can skip through any unexciting or uninteresting scenes but he or she would be dissatisfied if the rugby highlights were to miss any *try* events, hence we evaluate the recall rate for try events while also checking on the precision of highlight scenes for overall viewing experience. It is worth emphasizing that 'scoring events' are not the only highlights of a match; consumers may also want to see other non-scoring events which amount to human excitability. The duration of each highlight scene varies from about 10 to 20 seconds.

TABLE II
PERFORMANCE OF OUR APPROACH FOR DETECTING HIGHLIGHTS IN RUGBY BROADCASTS

| Matches | # of Try events | # of Try events detected in highlights | # of highlight scenes generated | # of scenes containing highlights* |
|---|---|---|---|---|
| Match 1 | 9 | 9 | 32 | 30 |
| Match 2 | 2 | 2 | 41 | 36 |
| Match 3 | 7 | 7 | 58 | 55 |
| Match 4 | 8 | 7 | 71 | 66 |
| Match 5 | 8 | 8 | 56 | 54 |
| **Total** | 34 | 33 | 258 | 241 |
| | **Try Recall** | 97.06% | **Highlights Precision** | 93.41% |

* A scene is considered as a highlight if it involves scoring events, non-scoring events like fouls or penalties, or events like foiling the defense or intercepting the ball from the opposition.

### C. User experience

To gauge the user experience, we asked 11 subjects (9 males, 2 females) to evaluate our highlights generator engine by giving a Mean Opinion Score (MOS), considering the following: i) the overall highlights viewing experience is enjoyable, entertaining and pleasant, and not marred by unexciting scenes; ii) the generated scenes do not begin or end abruptly; iii) the scenes are acoustically and/ or visually

TABLE III
MEAN OPINION SCORES

| | Match 1 | Match 2 | Match 3 | Match 4 | Match 5 |
|---|---|---|---|---|---|
| **MOS*** | 4.43 | 4.43 | 4.14 | 4.16 | 4.0 |

*MOS Rating: 1- Bad; 2- Poor; 3- Good; 4- Very Good; 5- Excellent

exciting. Each user watched at least ten scenes per match, picked at random. The overall MOS from Table III is calculated as 4.23 based on which we can state that the user experience is extremely positive.

## IV. DISCUSSION AND CONCLUSIONS

The precision error for 'whistle' category (see Table I) is a result of overlapping segments in the test data. Our experiments show that these segments under the whistle event category were sometimes classified as speech after the first classification. But using multi-stage GMM classification approach proved to be advantageous as all those segments were then recovered as 'excited speech'. The overlapping segments were rightly classified as such: a commentators' voice *is* in an excited state when a whistle is sounded. Further, we find that our engine failed to detect one 'try' in Match 4 (see Table II), and our analysis showed us that the excited speech was completely overshadowed by crowd sound for the most part of the decision window.

We also note that few studies in the literature have reported on actually generating sports highlights once key acoustic events are detected; our highlight generator engine effectively mines the exciting scenes. We add margins of a few seconds on either side of the generated highlight scenes to avoid abrupt beginning and ending of the highlight scenes, and doing so results in a positive experience for the users. Often live broadcasts are geographically targeted and hence the language of commentary may differ. Our language independent approach (using only excited speech and whistle sounds) easily addresses this concern.

In conclusion, we address the problem of generating highlights from rugby broadcasts using acoustic event detection. Our uni-modal approach makes use of a multi-stage classifier to detect key acoustic events which are fed to the highlight generator engine. The onset point and end-point selecting mechanism helps in setting the event boundaries and provides for *smoothness* and *completeness* of highlight viewing experience for the consumers. Given the simplicity and effectiveness of our highlights generator engine, it can be embedded in consumer electronics, and can be used for generating highlights in online settings (live TV broadcasts) as well as in offline scenarios (stored sports multimedia).